\documentclass[conference]{IEEEtran}
\IEEEoverridecommandlockouts
\usepackage{tipa}
\usepackage{balance}
\usepackage{epsfig}
\usepackage{amsmath}
\usepackage{theorem}
\usepackage{mathrsfs}
\usepackage{amsfonts}
\usepackage[utf8]{inputenc}
\usepackage{stackengine}
\def\delequal{\mathrel{\ensurestackMath{\stackon[1pt]{=}{\scriptstyle\Delta}}}}
\usepackage{tikz}
\usetikzlibrary{matrix,calc,arrows.meta,quotes,bending,positioning}

\DeclareMathOperator*{\argmin}{arg\,min}
\usepackage{pdfrender}
\usepackage{subfigure}
\usepackage{caption}
\usepackage{csquotes}
\newcommand*{\boldgreek}[1]{%
  \textpdfrender{%
    TextRenderingMode=FillStroke,%
    LineWidth=.35pt,%
  }{#1}%
}
\usepackage{subfigure}
\usepackage{tabularx}
\usepackage{caption}
\usepackage{csquotes}
\usepackage[font=footnotesize]{caption}
\usepackage{multirow}
\usepackage{amssymb}
\usepackage{mathrsfs}
\usepackage{euscript}
\usepackage{graphicx}
\usepackage{multirow}
\usepackage{xcolor}
\usepackage{algorithmicx}
\usepackage{algorithm}
\usepackage{algpseudocode}
\usepackage{cite}
\usepackage{url}
\usepackage{amsmath}
\usepackage[export]{adjustbox}
\usepackage{algorithm}
\usepackage{amssymb}
\usepackage{mathrsfs}
\usepackage{euscript}
\usepackage{graphicx}
\usepackage{multirow}
\usepackage{xcolor}
\usepackage{algorithmicx}
\usepackage{algorithm}
\usepackage{algpseudocode}
\usepackage{url}
\usepackage{amsmath}
\usepackage[export]{adjustbox}
\usepackage{algorithm}
\usepackage{diagbox}
\usepackage{epstopdf}
\begin{document}
\title{Xavier-Enabled Extreme Reservoir Machine for Millimeter-Wave Beamspace Channel Tracking}

\author{\IEEEauthorblockN {Hosein~Zarini$^{\dag}$, Mohammad Robat Mili$^{\S}$, Mehdi~Rasti$^{\dag, \star}$, Pedro H. J. Nardelli$^{\star, \star\star}$, and Mehdi Bennis$^{\star\star}$	
}
		$^{\dag}$Department of Computer Engineering, Amirkabir University of Technology, Tehran, Iran\\
		$^\S$Electronics Research Institute, Sharif University of Technology, Tehran, Iran\\
		$^\star$Lappeenranta-Lahti University of Technology, Lappeenranta, Finland\\
		$^{\star\star}$University of Oulu, Oulu, Finland\\
	}
\maketitle
\begin{abstract}
In this paper, we propose an accurate two-phase  millimeter-Wave (mmWave) beamspace channel tracking mechanism. Particularly in the first phase, we train an extreme reservoir machine (ERM) for tracking the historical features of the mmWave beamspace channel and predicting them in upcoming time steps. Towards a more accurate prediction, we further fine-tune the ERM by means of Xavier initializer technique, whereby the input weights in ERM are initially derived from a zero mean and finite variance Gaussian distribution, leading to 49\% degradation in prediction variance of the conventional ERM. The proposed method numerically improves the achievable spectral efficiency (SE) of the existing counterparts, by 13\%, when signal-to-noise-ratio (SNR) is 15dB. We further investigate an ensemble learning technique in the second phase by sequentially incorporating multiple ERMs to form an ensembled model, namely adaptive boosting (AdaBoost), which further reduces the prediction variance in conventional ERM by 56\%, and concludes in 21\% enhancement of achievable SE upon the existing schemes at SNR~=~15dB. 
\end{abstract}


\section{Introduction}

\par In recent years, beamspace multiple-input-multiple-output MIMO technology\cite{beamspace1} has been attended into a great extent, so as to compensate for the inefficient functionality of the conventional massive MIMO architecture in high-frequency bands, such as millimeter-Wave (mmWave) communications. In conventional mmWave massive MIMO systems, each antenna element requires a dedicated radio-frequency (RF) chain\footnote{RF chains are known as dominant modules in energy consumption, hardware cost and complexity order of conventional massive MIMO systems.}, which practically is prohibited. In contrast, the beamspace technology employs a large-scale antenna array mounted on the focal surface of an energy-focusing electromagnetic lens, whereby limited RF chains are concentrated on few dominant mmWave directions (beams), leading to a remarkable reduction in system dimensionality, hardware cost and energy consumption.
\par In a lens-aided architecture yet, the beamspace channel is very hard to be precisely estimated.
~In this regard, due to the severe computational burdens, the earlier channel estimation schemes e.g.,\cite{OMP}, based on complicated optimization techniques are ill-suited in current real-time applications. 
Thanks to their computational efficiency, deep learning techniques have been widely adopted for wireless network research over the recent years. For instance, the authors of \cite{LAMP2} and \cite{LAMP3}, applied deep learning techniques to overcome the computational burdens and improve the accuracy of the conventional approximate message passing (AMP)\cite{AMP} scheme for beamspace channel estimation. In vehicular communications as well, the authors of\cite{LSTM1} investigated the long-short-term-memory (LSTM) as a deep learning method for real-time channel estimation. In \cite{conf}, the authors exploited environmental samples to scrutinize the performance of the existing deep learning schemes, proposed for channel estimation. In specific, they investigated deep learning schemes in literature for predicting the beam angles on transceivers. The results nevertheless, revealed non-negligible errors, principally stemmed from training challenges such as overfitting in deep learning techniques.
\par Towards alleviating this drawback, we propose a precise two-phase mmWave beamspace channel estimation scheme in this paper from a time-series prediction perspective (i.e., a beamspace channel tracking scheme) as follows.
\begin{itemize}
    \item We utilize an extreme reservoir machine (ERM)\cite{RC} in the first phase for tracking the historical features of the mmWave beamspace channel as time-series sequences, and predict them in upcoming time steps, to construct the beamspace channel. Despite \cite{LAMP2}, \cite{LAMP3}, and \cite{LSTM1}, where a fully-connected structure is employed, ERM is advantageous for sparse connectivity, which thereby concludes lower training epochs to converge.
    \item The trained ERM is thereafter fine-tuned by means of Xavier initialization technique\cite{Xavier}. Through this way, the initial input weights in ERM are drawn from a finite variance and zero mean Gaussian distribution, which diminishes the prediction variance in conventional ERM by 49\%. In terms of achievable spectral efficiency (SE), the proposed fine-tuned ERM is being shown superior to the optimization-driven policy \cite{OMP} and the deep learning policy \cite{LSTM1}, by up to 27\% and 13\%, respectively, when the signal-to-noise-ratio (SNR) is 15dB.
    \item In the second phase, we additionally improve the performance of the proposed Xavier-enabled ERM scheme by sequentially incorporating multitude of ERMs (each one is known as a weak learner) into an ensembled model (known as a strong learner)\cite{L2Boost}. The proposed strong learner not only further advances the prediction accuracy of the weak learners each, but also alleviates their prediction variance by up to 56\%, while exhibiting the same complexity order as a weak learner. Numerically, the proposed ensembled model outperforms the achievable SE of the prior counterparts \cite{OMP} and \cite{LSTM1}, by up to 36\% and 21\%, respectively, when SNR~=~15dB.
\end{itemize}
\par The remainder of this paper is organized as follows. Section II and III describe the system setup and the problem statement, as well as the proposed solution. Finally, simulations and conclusions are presented in Sections IV and V, respectively.
\section{System Setup}
\begin{figure}
\centering\includegraphics[width=8.50cm,height=4.50cm]{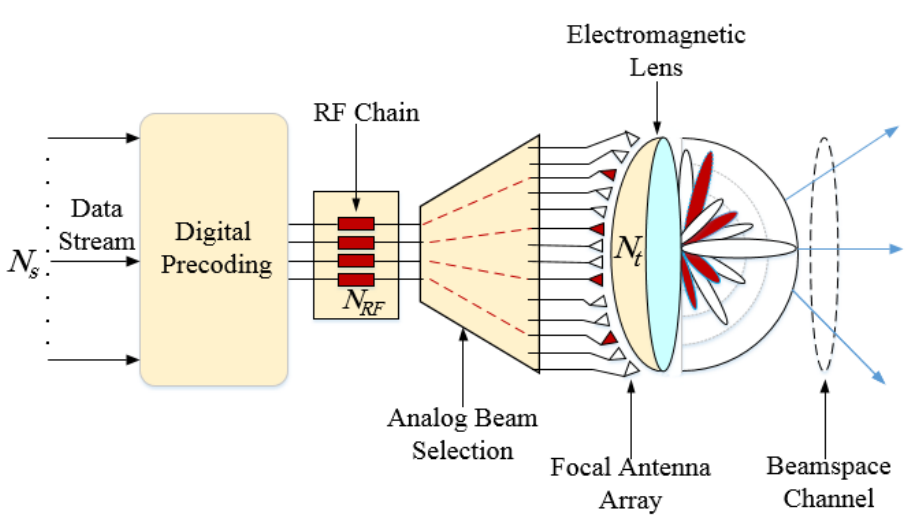}\caption{MmWave channel is transformed from the spatial domain into the beamspace domain using DFT operation\cite{wideband} in lens antenna array.}
\label{fig:lens}
\end{figure}
\subsection{Hybrid Analog-Digital Architecture}
We consider a downlink mmWave communication, wherein a transmitter employs $N_{\textrm{t}}$ ($N_{\textrm{t}}^{RF}$) transmit antennas (transmit RF chains) to
serve a receiver with $N_{\textrm{r}}$ ($N_{\textrm{r}}^{RF}$) receive antennas (receive RF chains) and the number of simultaneous data
streams (i.e., the system multiplexing gain) is $N_{\textrm{s}}$~=~min$(N_{\textrm{t}}^{RF},N_{\textrm{r}}^{RF})$. As a flexible, economical and energy efficient structure, we assume a hybrid analog-digital beamspace architecture\cite{beamspace1} for the transceivers.
According to Fig.~\ref{fig:lens}, the transmitter exploits a baseband digital matrix $\mathbf{F}_{\textrm{BB}}\in
\mathbb{C}^{N_{\textrm{t}}^{RF}\times N_{\textrm{s}}}$ for precoding the normalized-power transmit symbols $\mathbf{s}\in
\mathbb{C}^{N_{\textrm{s}}\times1}$, where $\mathbb{E}\left[  \mathbf{s}\mathbf{s}^{H}\right]  =\mathbf{I}_{N_{\textrm{s}}}$. Subsequently, the analog beam selection
network at the transmitter $\mathbf{S}_{\textrm{t}}\in{\mathbb{R}^{N_{\textrm{t}}\times{N_{\textrm{t}}^{RF}%
}}}$ maps $N_{\textrm{t}}^{RF}$ RF chains into a subset of $N_{\textrm{t}}$ transmit antennas, using analog
switches. An energy-focusing electromagnetic lens is afterwards embedded with a deployed antenna array on its focal surface. At the receiver side reversely, the lens antenna array receives the signals, where the analog beam selection network $\mathbf{S}_{\textrm{r}}$ $\in \mathbb{R}^{N_{\textrm{r}}\times N_{\textrm{r}}^{RF}}$ maps a subset of the receive antennas into the RF chains. As the transmitter, a digital baseband combining matrix $\mathbf{W}_{\textrm{BB}}$ $\in \mathbb{C}^{N_{\textrm{r}}^{RF}\times
N_{\textrm{s}}}$ is responsible at the receiver to obtain the symbols. Hence, the received
discrete-time complex baseband signal $y$ can be expressed as
$y=\mathbf{W}_{\textrm{BB}}^{H}\mathbf{S}_{\textrm{r}}^{H}\mathbf{H}_{b}\mathbf{x}+\mathbf{W}%
_{\textrm{BB}}^{H}\mathbf{S}_{\textrm{r}}^{H}\mathbf{n}$,
where $\mathbf{H}_{b}$ is the mmWave beamspace channel generated by the discrete Fourier transformation (DFT) in lens antenna array\cite{wideband} and $\mathbf{n}$ $\sim N(0,\sigma^{2}\mathbf{I}_{N_{\textrm{r}}})$ denotes the
additive white Gaussian noise (AWGN) with $\sigma^{2}$, denoting the noise power.
\subsection{Communicating MmWave Channel} 
Consider a clustered ray-based mmWave channel based on the well-known
Saleh-Valenzuela geometric model \cite{Saleh}, with $N_{{cl}}$ cluster of scatterers, where a typical cluster $l$ has a limited angle-of-departure/arrival (AoD/AoA) spread, denoted
by $\psi_{\textrm{t}}^{l}$ and $\psi_{\textrm{r}}^{l}$ respectively. It is assumed that the $l$th cluster is contributed with $N_{{ray}}$ propagation rays, where a typical ray $u$, has a physical AoD/AoA denoted by $\theta_{\textrm{t}}^{{l,u}}\in \psi_{\textrm{t}}^{{l}}$ and $\theta_{\textrm{r}}^{{l,u}}\in \psi_{\textrm{r}}^{{l}}$, respectively, as well as a complex-valued gain, denoted by $\alpha^{{l,u}}$.
Let us denote, the spatial AoD/AoA, by $
\phi_{\textrm{t}}^{{l,u}}=(d/\lambda)\sin \theta_{\textrm{t}}^{{l,u}}$ and $\phi
_{\textrm{r}}^{{l,u}}=(d/\lambda)\sin \theta _{\textrm{r}}^{{l,u}},$ respectively, in which $d$ denotes the
antenna element spacing and $\lambda$ is the wavelength. Hence, the discrete-time narrowband mmWave channel $\mathbf{H}%
\in \mathbb{C}^{N_{\textrm{r}}\times N_{\textrm{t}}}$  is given by $\mathbf{H}=\gamma\sum_{{l}=1}^{N_{{cl}}} \sum_{{u}=1}^{N_{{ray}}} \alpha ^{{l,u}}\mathbf{a}_{\textrm{r}}\left( \phi _{\textrm{r}}^{{l,u}}\right)\mathbf{a}%
_{\textrm{t}}^{\textrm{H}}\left( \phi _{\textrm{t}}^{{l,u}}\right),$
where $\gamma$ is the normalization factor given by $\gamma=\sqrt{N_{\textrm{r}}N_{\textrm{t}}/{N_{{cl}}N_{{ray}}}}.$
Also, $\mathbf{a}_{\textrm{t}}\left( \phi _{\textrm{t}}^{{l,u}}\right) \in \mathbb{C}^{N_{\textrm{t}}\times 1}$
and $\mathbf{a}_{\textrm{r}}\left( \phi _{\textrm{r}}^{{l,u}}\right) \in \mathbb{C}^{N_{\textrm{r}}\times 1}$,
denote the antenna array responses at the transmitter and receiver, respectively, based on uniform linear array (ULA) and given by $\mathbf{a}_{\textrm{t}}\left( \phi_{\textrm{t}}^{{l,u}}\right)  =\frac{1}{\sqrt{N_{\textrm{t}}}
}\left[ 1,e^{j2\pi \phi_{\textrm{t}}^{{l,u}}},...,e^{j2\pi \left( N_{\textrm{t}}-1\right)
\phi_{\textrm{t}}^{{l,u}}}\right] ^{H}$ and $\mathbf{a}_{\textrm{r}}\left( \phi_{\textrm{r}}^{{l,u}}\right)  =\frac{1}{\sqrt{N_{\textrm{r}}}}
\left[ 1,e^{j2\pi \phi_{\textrm{r}}^{{l,u}}},...,e^{j2\pi \left( N_{\textrm{r}}-1\right)
\phi_{\textrm{r}}^{{l,u}}}\right] ^{H}$. In this regard relying on DFT operations\cite{wideband} in lens antenna array, the mmWave channel $\mathbf{H}$ in spatial domain can be efficiently transformed into the
equivalent channel in beamspace domain i.e., $\mathbf{H}_{b}$ (see \cite{wideband}).

\subsection{Problem Statement}
\par For given the transceiver design parameters $\mathbf{F}_{\textrm{BB}}, \mathbf{W}_{\textrm{BB}}$, ${\mathbf{S}_{{t}}}$ and ${\mathbf{S}_{{r}}}$, the beamspace channel estimation as a signal recovery optimization problem\cite{LAMP2} is defined as
\begin{align}\label{opt2}
    &\min_{\mathbf{H}_{b}}~||\mathbf{H}_{b}||_{0}^{2}
    \\&\nonumber s.t. 
    ~||y-\sqrt{\psi /N_{\textrm{s}}}\mathbf{W}_{\textrm{BB}}^{\textrm{H}}\mathbf{S}_{\textrm{r}}^{\textrm{H}}\mathbf{H}_{{b}}\mathbf{S}_{\textrm{r}}\mathbf{F}_{\textrm{BB}}\mathbf{s}||\le{\varepsilon},
\end{align}
where $||.||_{0}$ indicates the number of non-zero elements and $\varepsilon$ is the error tolerance. From a time-series prediction perspective, we are seeking for minimizing $||\mathbf{H}_b(t)-\widehat{\mathbf{H}_b}(t)||_{2}$, where $\widehat{\mathbf{H}_b}(t)$ is the predicted beamspace channel at time step $t$. This is achieved via a two-phase mmWave beamspace channel tracking mechanism, proposed in this paper as follows.
\section{First phase: MmWave Beamspace Channel Tracking via ERM}
In the first phase, ERM is briefly introduced and fine-tuned to learn the historical mmWave beamspace channel features, and forecast them in upcoming timesteps.
\subsection{Training Sample Set Acquisition}
We define a motion feature state vector for a typical transceiver as
$\boldgreek{\chi} \delequal [\boldgreek{{\theta}_\textrm{r}},\boldgreek{\theta_\textrm{t}}]$, with $\boldgreek{\theta}_\textrm{r}=\{\theta_\textrm{r}^1,\theta_\textrm{r}^2,...,\theta_\textrm{r}^{|N^{eff}_{\textrm{r}}|}\}$ and $\boldgreek{\theta}_\textrm{t}=\{\theta_\textrm{t}^1,\theta_\textrm{t}^2,...,\theta_\textrm{t}^{|N^{eff}_{\textrm{t}}|}\}$, where  $|N^{eff}_{\textrm{t}}|=|N_{\textrm{t}}|\times|N_{cl}|\times|N_{ray}|$, $|N^{eff}_{\textrm{r}}|=|N_{\textrm{r}}|\times|N_{cl}|\times|N_{ray}|$, and $|N^{eff}|=|N^{eff}_{\textrm{t}}|+|N^{eff}_{\textrm{r}}|$. 
As in \cite{LSTM1}, we generally assume that the feature states such as AoDs and AoAs are time-varying and follow their historical time steps i.e., $\boldgreek{\chi}(1),\boldgreek{\chi}(2),...,\boldgreek{\chi}(t-1)$, indicating a time-series form to predict $\boldgreek{\chi}(t)$. Relying on the predicted values in $\boldgreek{\chi}(t)$ afterwards, one can derive the non-zero elements in beamspace channel $\textbf{H}_b$ through the pilot transmission based on the classical least square method\cite{LAMP2}. In what follows, it is well elaborated how to predict $\boldgreek{\chi}(t)$, relying on its historical time-series $\boldgreek{\chi}(1),\boldgreek{\chi}(2),...,\boldgreek{\chi}(t-1)$.
\subsection{Time-Series Forecasting}
Artificial neural networks (ANN) with temporal dynamic behavior and time-varying nature, are renowned for their time-series data (i.e., sequences of data over the time steps) processing capability. To this goal, time-series ANN are trained by exploiting the historical knowledge and to predict them in next time steps. The feature states of the transceivers generally exhibit dynamic characteristics over the time\cite{conf} and thus, time-series ANNs are leveraged in this paper for predicting their upcoming time steps.
\par
\begin{figure}
\centering\includegraphics[width=8.50cm,height=4.50cm]{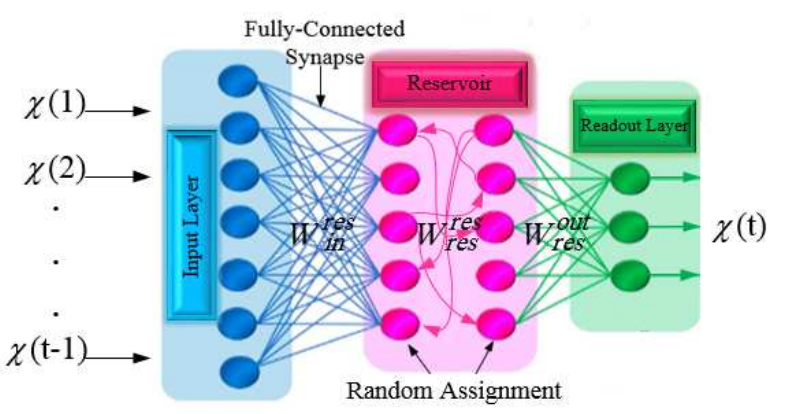}
\caption{The architecture of ERM.}
\label{fig:ERM}
\end{figure}

 As a time-series ANN, ERM is a reservoir computing\cite{RC} trained by an extreme learning policy\cite{ex}, whereby the sparse neurons in its reservoir layer(s) are randomly associated with each other (rather than the fully-connected structure) and remain untuned during the training process. Thanks to the random and sparse connectivity of neurons in reservoir layer(s), ERM is superior to the conventional fully-connected counterparts in terms of prediction time. 
As depicted in Fig.~\ref{fig:ERM}, the input layer of ERM is fed by the past motion state features $\boldgreek{\chi}(1),\boldgreek{\chi}(2),...,\boldgreek{\chi}(t-1)$, and the reservoir layer state in time step $t$ is given by\cite{RC}
\begin{align} \label{RC}
    Res(t) = f^{act}[\widehat{W}_{res}^{res}.Res(t-1)+\widehat{W}_{in}^{res}.{\boldgreek{\chi}}(t-1)],
\end{align}
with $\widehat{W}_{res}^{res}$ and $\widehat{W}_{in}^{res}$, indicating the weight connections of reservoir layer(s) and input layer, respectively. The hidden weights on reservoir layer $\widehat{W}_{res}^{res}$ as already mentioned, are randomly assigned and never updated (i.e. they follow a random projection with non-linear transforms), or can be inherited from their ancestors (if multitude of reservoir layers coexist) without being changed.
The activation function $f^{act}$ in (\ref{RC}), is a sigmoid or logsig function, which injects non-linearity to the training process. 
The readout layer according to Fig.~\ref{fig:ERM}, predicts the motion state features at time step $t$ as following
\begin{align}
    \boldgreek{\chi}(t) = \widehat{W}_{res}^{out}.Res(t-1),
\end{align}
with $\widehat{W}_{res}^{out}$, denoting the weights of readout layer in ERM and mostly learned in a single step, which essentially amounts to learning a linear model\cite{RC}. 
\par
\begin{figure}
\centering\includegraphics[width=8.50cm,height=4.5cm]{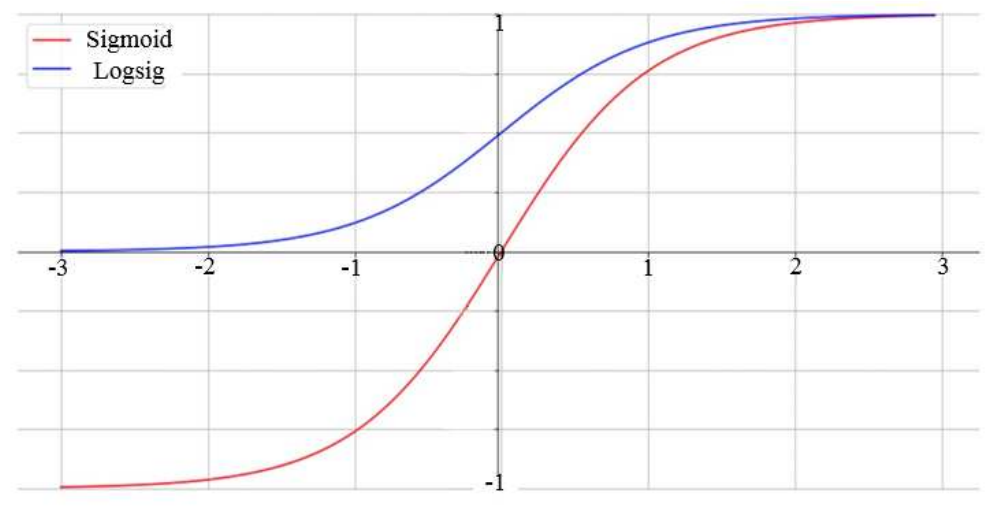}
\caption{The behaviour of sigmoid and logsig non-linear functions.}
\label{fig:sigmoid-logsig}
\end{figure}

 In (\ref{RC}), initializing the input weights $\widehat{W}_{in}^{res}$ by proper values is crucial for ERM to converge within a reasonable amount of iterations/epochs. The behaviour of activation functions clarifies this requirement more.
One can observe from Fig.~\ref{fig:sigmoid-logsig} that both functions expose an approximately linear behaviour for the input values adjacent to zero. If the input weights in (\ref{RC}) are initialized too small, the variance of the input signals $\boldgreek{\chi}(1),\boldgreek{\chi}(2),...,\boldgreek{\chi}(t-1)$ starts diminishing layer by layer, ending up with trivial values which basically removes the non-linearity feature of $f^{act}$ and thereby, the advantage of deep layers is lost. On the opposite side, both of the functions in Fig.~\ref{fig:sigmoid-logsig}, tend to become flat and saturate for larger values of input signals. By adopting the initial weights too large, the variance of the input signals tends to escalate in a layer-wise fashion. In this case, the gradient starts approaching zero, thereby updating the weights will not be properly accomplished. In the following section, we provide a brief overview of the Xavier initializer to regulate the input signal initial weights.
\subsection{Xavier Initializer}
Towards a proper weight initialization, we answer the following questions on the basis of Xavier Initialization technique. 1) Which distribution is adequately proper for $\widehat{W}_{in}^{res}$ to be drawn from. 2) What is the effective variance of $\widehat{W}_{in}^{res}$ to this end.
Suppose that the well-known Gaussian distribution with a zero mean and a finite variance is the candidate. For the sake of exposition, consider (\ref{RC}) for a typical user in a time step, which is a linear combination of the input element $\chi_j$ and its corresponding weight $\widehat{w}_{in}^{res}(j)\in{\widehat{W}_{in}^{res}}$ as $y^{lin}=\widehat{w}_{in}^{res}(1).\chi_{1}+\widehat{w}_{in}^{res}(2).\chi_2+...+\widehat{w}_{in}^{res}({|N^{eff}|}).\chi_{|N^{eff}|}$.
The aim here is to keep the variance of $\chi_j$ and $y^{lin}$ identical. The variance of $y^{lin}$ can be given by
\begin{align} \label{var1}
    \textrm{var}(y^{lin}) \!=& \textrm{var}\big(\widehat{w}_{in}^{res}(1).\chi_{1}+\\&\nonumber\widehat{w}_{in}^{res}(2).\chi_2+...+\widehat{w}_{in}^{res}({|N^{eff}|}).\chi_{|N^{eff}|}\big).
\end{align}
By definition of variance, it can be concluded that
\begin{align} \label{var2}
    \textrm{var}(\widehat{w}_{in}^{res}(j).\chi_{j}) \!=& \mathbb{E}(\widehat{w}_{in}^{res}(j))^2.\textrm{var}(\chi_{j}) +\\&\nonumber \mathbb{E}(\chi_{j})^2.\textrm{var}(\widehat{w}_{in}^{res}(j)) \!\!+\!\! \textrm{var}(\widehat{w}_{in}^{res}(j)).\textrm{var}(\!\chi_{j}\!),
\end{align}
where $\mathbb{E}(.)$ stands for the expectation or mean value. For the Gaussian distribution with zero mean, (\ref{var2}) will be given by $\textrm{var}(\widehat{w}_{in}^{res}(j).\chi_{j}) = \textrm{var}(\widehat{w}_{in}^{res}(j)).\textrm{var}(\chi_{j})$.
Therefore, (\ref{var1}) can be rewritten as follows
\begin{align} \label{var3}
    \textrm{var}(y^{lin}) =& \textrm{var}(\widehat{w}_{in}^{res}(1)).\textrm{var}(\chi_{1}) +\textrm{var}(\widehat{w}_{in}^{res}(2)).\textrm{var}(\chi_{2})+\\&\nonumber ... +\textrm{var}(\widehat{w}_{in}^{res}(|N^{eff}|)).\textrm{var}(\chi_{|N^{eff}|}).
\end{align}
For the identical distribution of weights and input elements, (\ref{var3}) can be stated as $\textrm{var}(y^{lin}) = |N^{eff}|.\textrm{var}(\widehat{w}_{in}^{res}(j)).\textrm{var}(\chi_{j})$.
So, if we want the variance of $\chi_j$ and $y^{lin}$ being identical, we can conclude that $|N^{eff}|.\textrm{var}(\widehat{w}_{in}^{res}(j)) = 1$ and $\textrm{var}(\widehat{w}_{in}^{res}(j)) = 1/|N^{eff}|$, 
which verifies that the weights should be drawn from a Gaussian distribution with zero mean and a finite variance of $1/|N^{eff}|$.
\begin{figure*}
\centering\includegraphics[width=12.0cm,height=6.50cm]{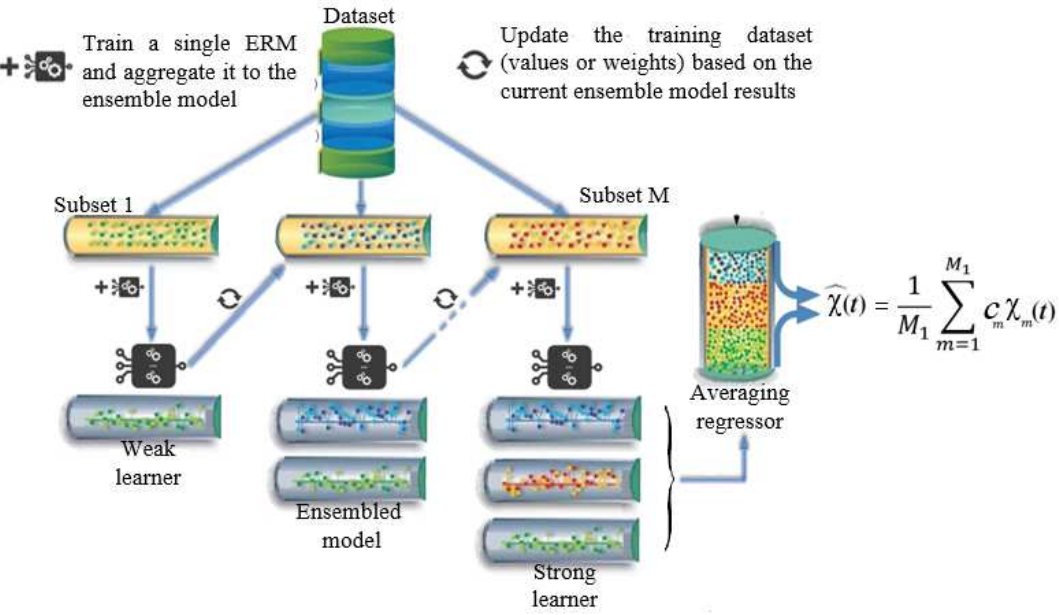}
\caption{Adaptive boosting mechanism schematic.}
\label{fig:ensemble}
\end{figure*}
\subsection{Second Phase: Enhancing Accuracy via Ensemble Learning}
In the second phase, adaptive boosting (AdaBoost) is briefly reviewed and investigated as an effective strategy of ensemble learning, so as to improve the prediction capability of ERM in mmWave beamspace channel tracking. The idea in ensemble learning is to train a strong ensembled learner, which merges the predictions made by various weak learners (e.g., the Xavier-enabled ERM modules in this paper) to make a more accurate prediction \cite{L2Boost}. In doing so, we adopt an AdaBoost mechanism according to Fig.~\ref{fig:ensemble}, whereby the weak learners are sequentially trained in an adaptive fashion. In fact, it is aimed at boosting the performance of the current weak learner through focusing on the improperly trained observations by the prior weak learners.
Since the ERMs predict on continuous domain, the beamspace channel tracking is a regression problem and the ERMs are regressors, as well. First, $M_1$ subsets $\chi_m (m\in M_1)$  are allocated to $M_1$ ERM regressors as the weak learners, which are sequentially trained for predicting the next time step, i.e., $\chi (t+1)$. The strong ensembled learner is defined based on a weighted sum of $M_1$ weak learners as $\Phi_{M_{1}}^{ens}=\sum_{m=1}^{M_1}c_{m}\chi_{m}$, with $c_{m}$ denoting the weight of the $m$th ERM as the performances of this weak model, i.e., the better a weak learner performs, the more it contributes to the strong ensembled model. The strong ensembled learner thus, is generally less biased than the weak learners. The main issue is to achieve the optimal order for the ERMs to be deployed along the ensembling chain. Note that the optimal $\Phi_{M_{1}}^{ens}$ is complicated to obtain when the ensembling chain is long. Instead, rather than globally seeking among the weak learners all, the best possible pairs of ($c_{m},\chi_{m}$) are locally built and added to the strong ensembled model one by one, in an iterative sub-optimal manner. Recurrently, the strong ensembled model can be indicated by $\Phi_{m}^{ens}=\Phi_{m-1}^{ens}+c_{m}\chi_{m}$, and the best ($c_{m},\chi_{m}$) pair is acquired as $(c_{m},\chi_{m}) =\argmin_{c,\chi}E(\Phi_{m-1}^{ens}+c\chi)$, where $E(.)$ indicates the fitting error of the strong ensembled model. The prediction made by the strong ensembled model by contributing the weak learners and their impact at time step $t$ finally, is given by $\hat{\chi}(t)=\frac{1}{M_{1}}\sum_{m=1}^{M_{1}}c_{m}\chi_{m}(t)$.
\subsection{Complexity Analysis}
The computational complexity for training ERM mainly incorporates the training time for its readout weights ${W}_{res}^{out}$,
evaluated as $\mathcal{O}(\textrm{ERM})=(T_{\text{max}}-T_0+1)^3$, with $T_{\text{max}}$ and $T_0$ indicating the maximum training time and the initial washout time, respectively\cite{RC1}. In an AdaBoost ensembled model nonetheless, since multiple ERM modules are sequentially incorporated, the complexity can be expressed as $\mathcal{O}(\textrm{ENS})=\mathcal{O}(\textrm{ERM}_{1})+\mathcal{O}(\textrm{ERM}_{2})+...+\mathcal{O}(\textrm{ERM}_{M_{1}})={max}\{\mathcal{O}(\textrm{ERM}_{j})\},~j\in{\{1,2,...,M_{1}\}}$.

\begin{table}
\centering
\captionsetup{font=small}
\captionsetup{justification=centering}
\caption{\\ERM configurations}
\label{table:notations}
\begin{tabular}
{ | m{3.0cm} | m{2cm} | }
\hline
\textbf{ERM Parameter} & \textbf{Value} \\ 
\hline\hline
TrainingSize & 75\% \\
\hline
ValidationSize & 25\% \\
\hline
Number of batches & 250 \\
\hline
No. input delay $n_{\boldgreek{\theta}_{\textrm{t}}}$ & 6\\
\hline
No. output delay $n_{\boldgreek{\theta}_{\textrm{r}}}$ & 6 \\
\hline
No. observations & 20 \\
\hline
Observation interval & 1(s) \\
\hline
\end{tabular}
\end{table} 
\section{Simulation Results}
We consider a clustered mmWave channel model incorporating $N_{{cl}}=8$ clusters, where each cluster contributes $N_{{ray}}=6$ propagation rays. The signal wavelength is supposed to be $\lambda$ = 1.36, while AoAs/AoDs are uniformly distributed within $[-\frac{1}{2},\frac{1}{2}]$ and the complex path-gain follows $\mathcal{CN}$(0, 1). Simulations are performed in a lens-aided MIMO system equipped with $N_{\textrm{t}}=64, N_{\textrm{r}}=16$ and $N_{\textrm{s}}=N_{\textrm{t}}^{\textrm{RF}}=N_{\textrm{r}}^{\textrm{RF}}=4$. As observed from Table I, 10000 samples are investigated in 250 batches, contributing 75\% for the training and the rest for the validation and test, where we observe 20 time slots of tracking with 1sec interval and the number of delay steps for the input and output layers are $n_{\theta_{\textrm{t}}}$~=~$n_{\theta_{\textrm{r}}}$~=~6.
\subsection{Performance Measures}
To evaluate the performance of our proposed method, we analyze the error metrics, RMSE and NMSE over $Q$ samples and respectively given by $\textrm{RMSE} = \sqrt{\sum_{i=1}^{{Q}}||\mathbf{H}^{i}_b(t)-\widehat{\mathbf{H}^{i}_b}(t)||^{2}_{2}}$,
and $\textrm{NMSE} = \mathbb{E}\Bigg\{\sum_{i=1}^{{Q}}||\mathbf{H}^{i}_b(t)-\widehat{\mathbf{H}^{i}_b}(t)||^{2}_{2}\times\big(\sum_{i=1}^{{Q}}||\widehat{\mathbf{H}^{i}_b}(t)||^{2}_{2}\big)^{-1}\Bigg\}$,
where ${\mathbf{H}^{i}_b}(t)$ and $\widehat{\mathbf{H}^{i}_b}(t)$ are the ground truth and the predicted values of the mmWave beamspace channel for the $i$th sample on time step $t$, respectively. 
We also define the standard deviation and the mean absolute deviation metrics for evaluating the beamspace channel tracking variance, 
respectively given by $\tau_1 = \sqrt{\sum_{i=1}^{{M}}||\widehat{\mathbf{H}^{i}_b}(t)-\mu||^{2}_{2}\times Q^{-1}}$,
and $\tau_2 = \sum_{i=1}^{{M}}||\widehat{\mathbf{H}^{i}_b}(t)-\mu||_{2}\times Q^{-1}$, with $M$ and $\mu$, denoting the number of predictions and the prediction average, respectively.
Finally, the achievable SE is expressed as
$\!SE \!=  \textrm{log}_{2}\big|\textbf{I}_{N_{\textrm{s}}}\!+\!\frac{\rho}{\sigma^{2}N_{\textrm{s}}}R_{n}^{-1}(\textbf{W}_{\textrm{BB}})^{H}(\textbf{S}_{\textrm{r}})^{H}\textbf{H}_{{b}}\textbf{S}_{\textrm{t}}\textbf{F}_{\textrm{BB}} (\textbf{F}_{\textrm{BB}})^{H}(\textbf{S}_{\textrm{t}})^{H}(\textbf{H}_{{b}})^{H}\textbf{S}_{\textrm{r}}\textbf{W}_{\textrm{BB}}\big|,$
where $R_{n}=\!(\textbf{W}_{\textrm{BB}})^{H}(\textbf{S}_{\textrm{r}})^{H}\textbf{S}_{\textrm{r}}\textbf{W}_{\textrm{BB}}$ is the noise covariance matrix after combining.
\begin{figure}
\centering\includegraphics[width=6.0cm,height=4.50cm]{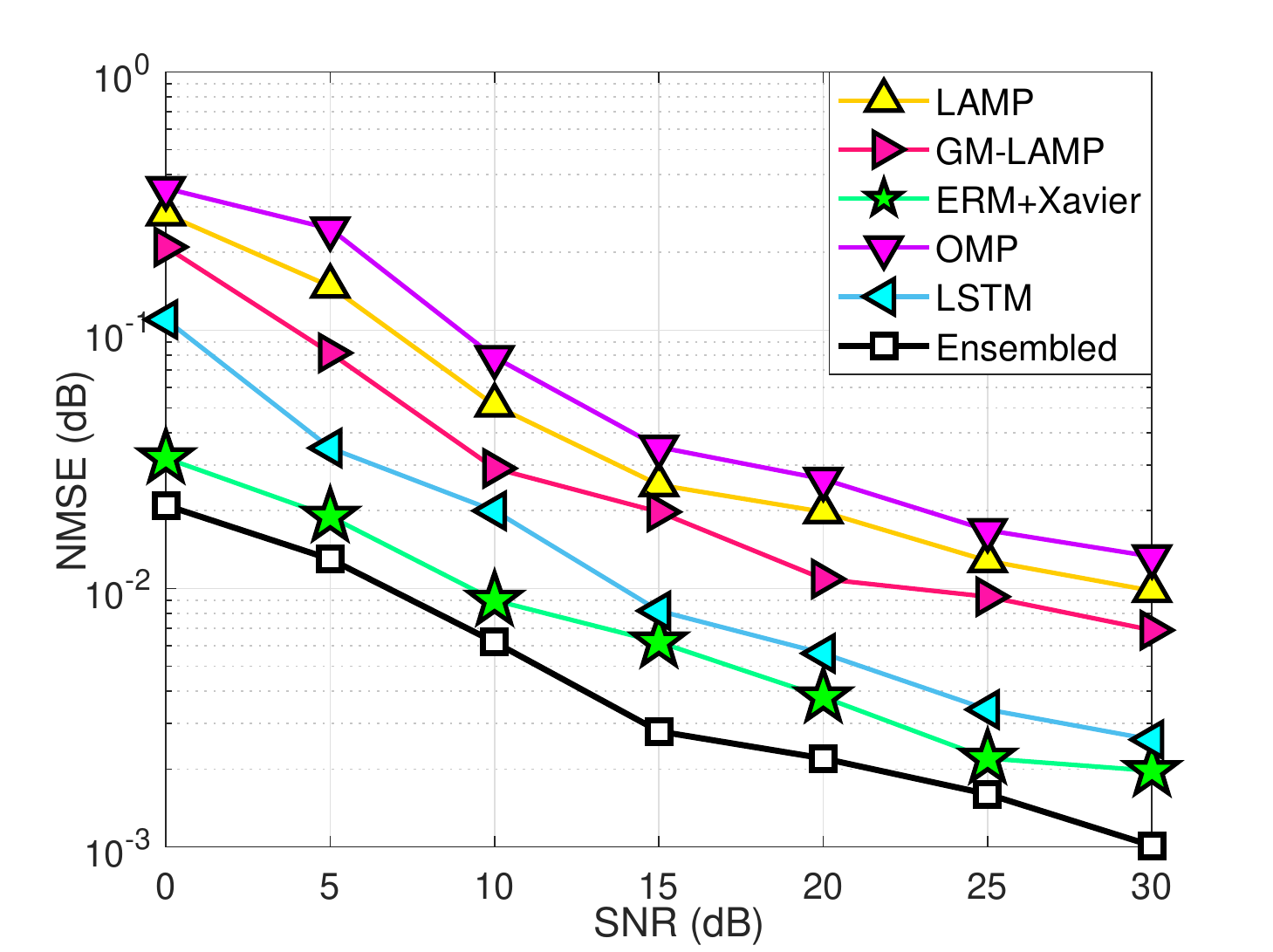}\caption{NMSE versus the SNR. Our proposed ERM+Xavier scheme outperforms LSTM and other baseline schemes especially in low SNR values.}
\label{fig:nmse}
\end{figure}
\begin{figure}
\centering\includegraphics[width=6.0cm,height=4.50cm]{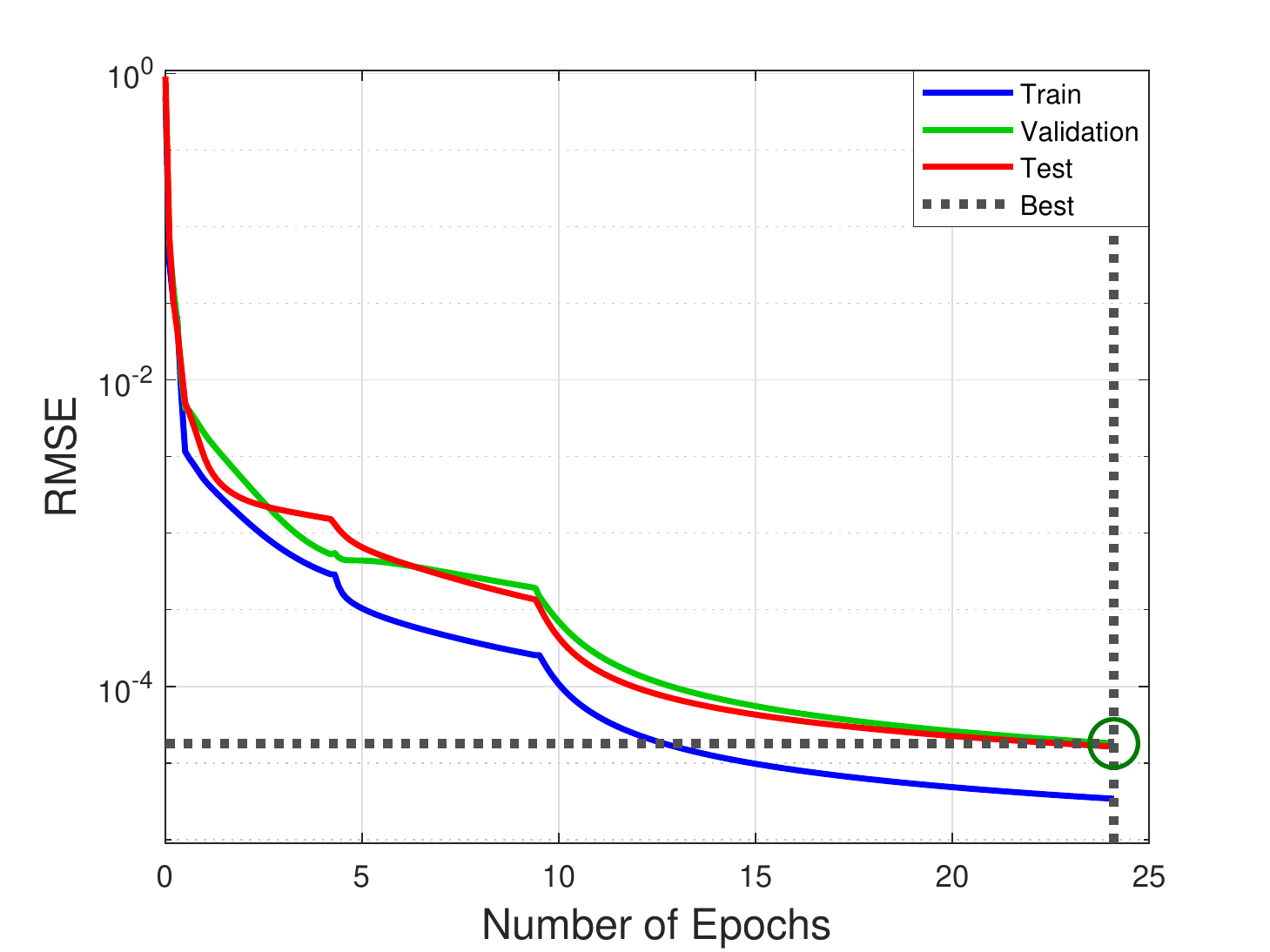}\caption{RMSE versus the number of convergence epochs.}
\label{fig:rmse}
\end{figure}

\subsection{Performance Evaluation}
We use Gaussian mixture learning-aided AMP (GM-LAMP)\cite{LAMP2}, learning-aided AMP (LAMP)\cite{LAMP3}, LSTM\cite{LSTM1} and orthogonal matching pursuit (OMP)\cite{OMP}, as the baseline schemes to evaluate the efficiency of the proposed ERM+Xavier and ensembeld schemes. We have trained the LAMP and GM-LAMP schemes under the simulation parameters given in \cite{LAMP2}, while the configurations for LSTM and ERM are identical. 
\begin{figure}
\centering\includegraphics[width=6.0cm,height=4.50cm]{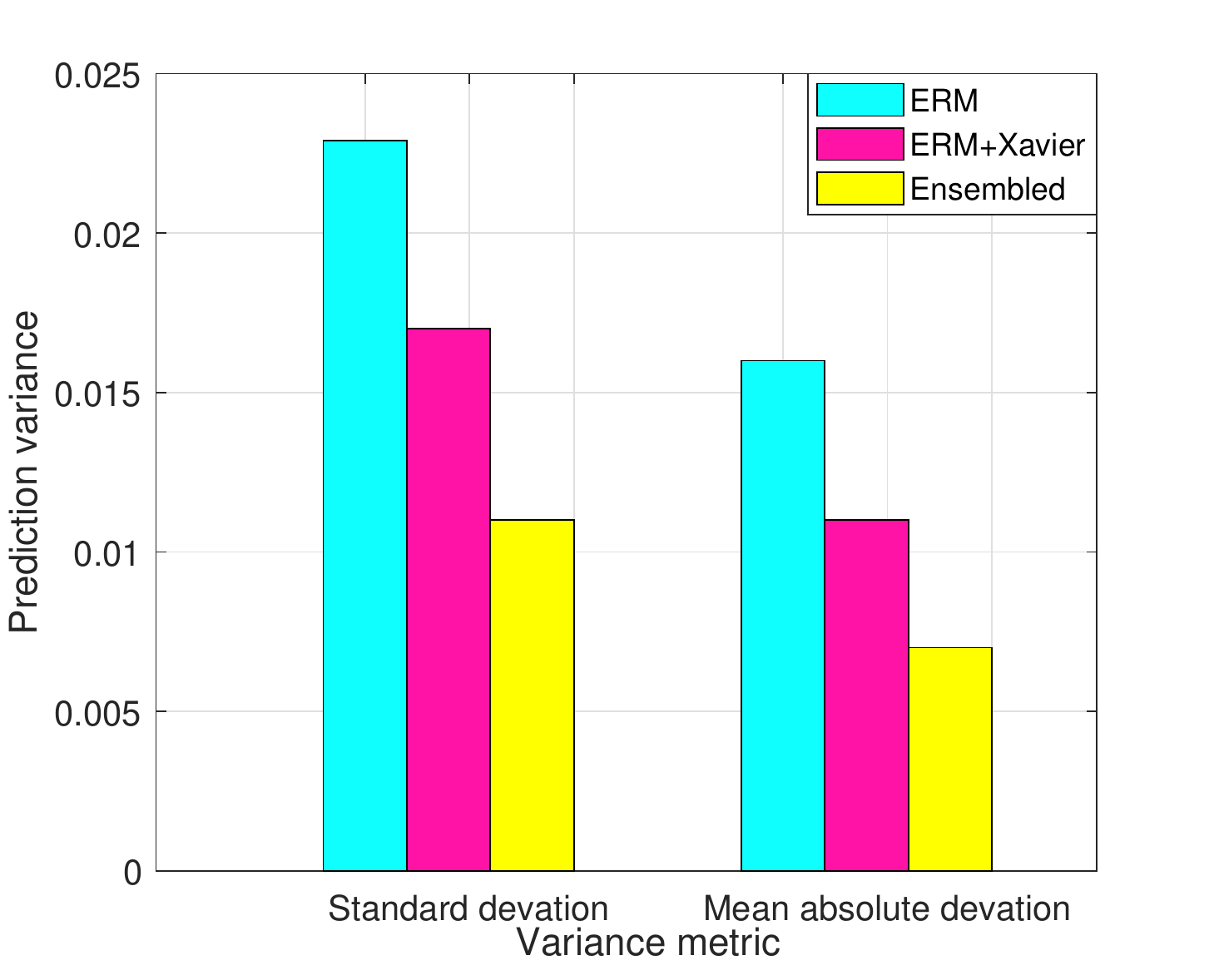}\caption{Analyzing the prediction variance by mean absolute deviation and standard deviation.}
\label{fig:var}
\end{figure}

\begin{figure}\vspace{-1.30 cm}
\centering\includegraphics[width=8.0cm,height=6.50cm]{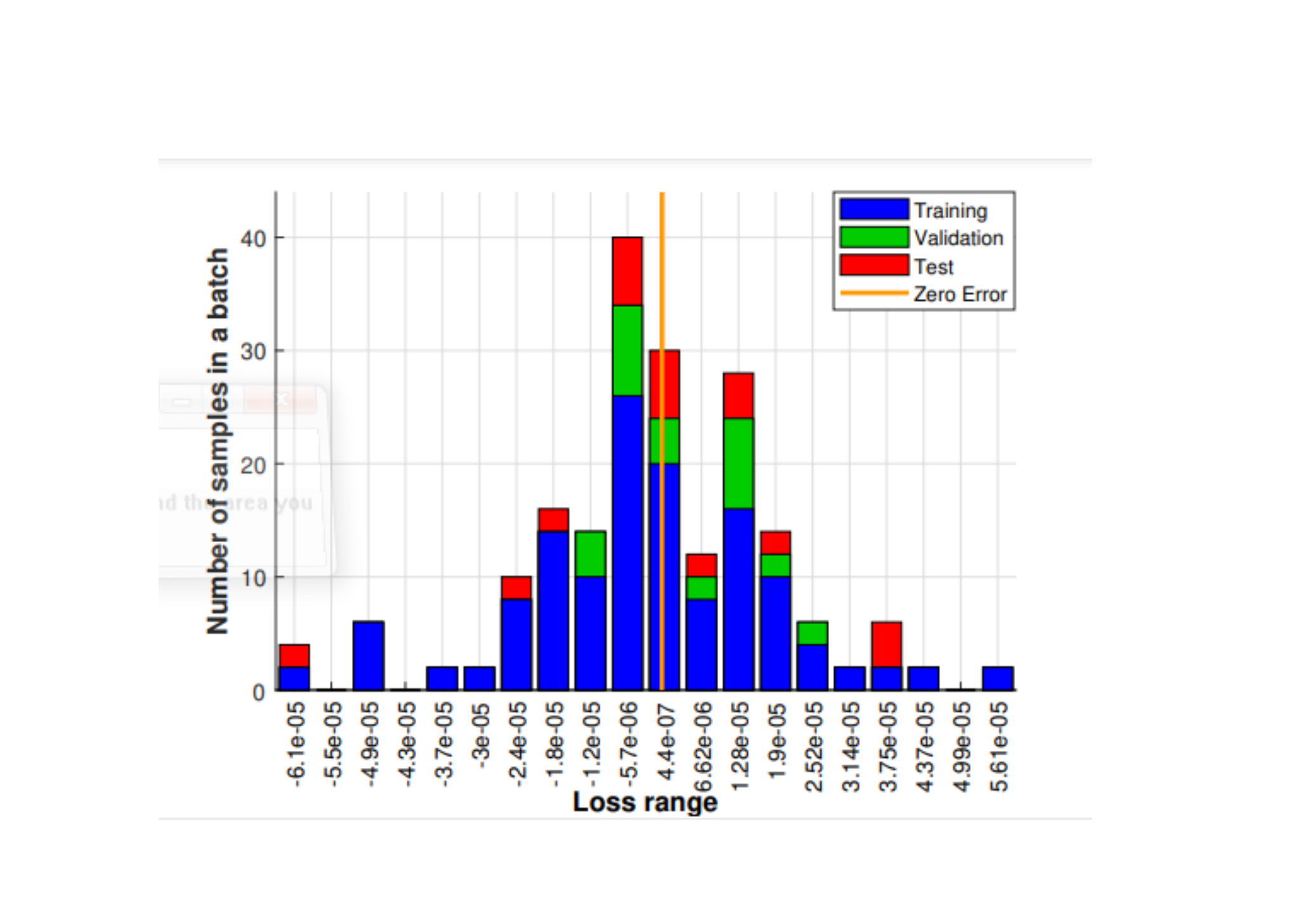}\caption{Number of samples vs. loss range.}
\label{fig:loss}
\end{figure}
\par In Fig.~\ref{fig:nmse} for varying SNR in 0dB$\sim$30dB, we analyze the NMSE of the baseline schemes. By increasing SNR, all the schemes evidently perform more accurately, leading to a degradation trend for the NMSE. As observed, our proposed ERM+Xavier scheme as a fine-tuned deep learning technique outperforms others, due to the higher time-series prediction accuracy. Furthermore, the ensembled model exhibits lower NMSE compared to all the schemes, thanks to a lower prediction variance, as well as an enhanced prediction capability.
Fig.~\ref{fig:rmse} indicates ERM convergence behaviour during training, validation and test procedure with negligible RMSE at the convergence
point. 
\begin{figure}
\centering\includegraphics[width=6.0cm,height=4.50cm]{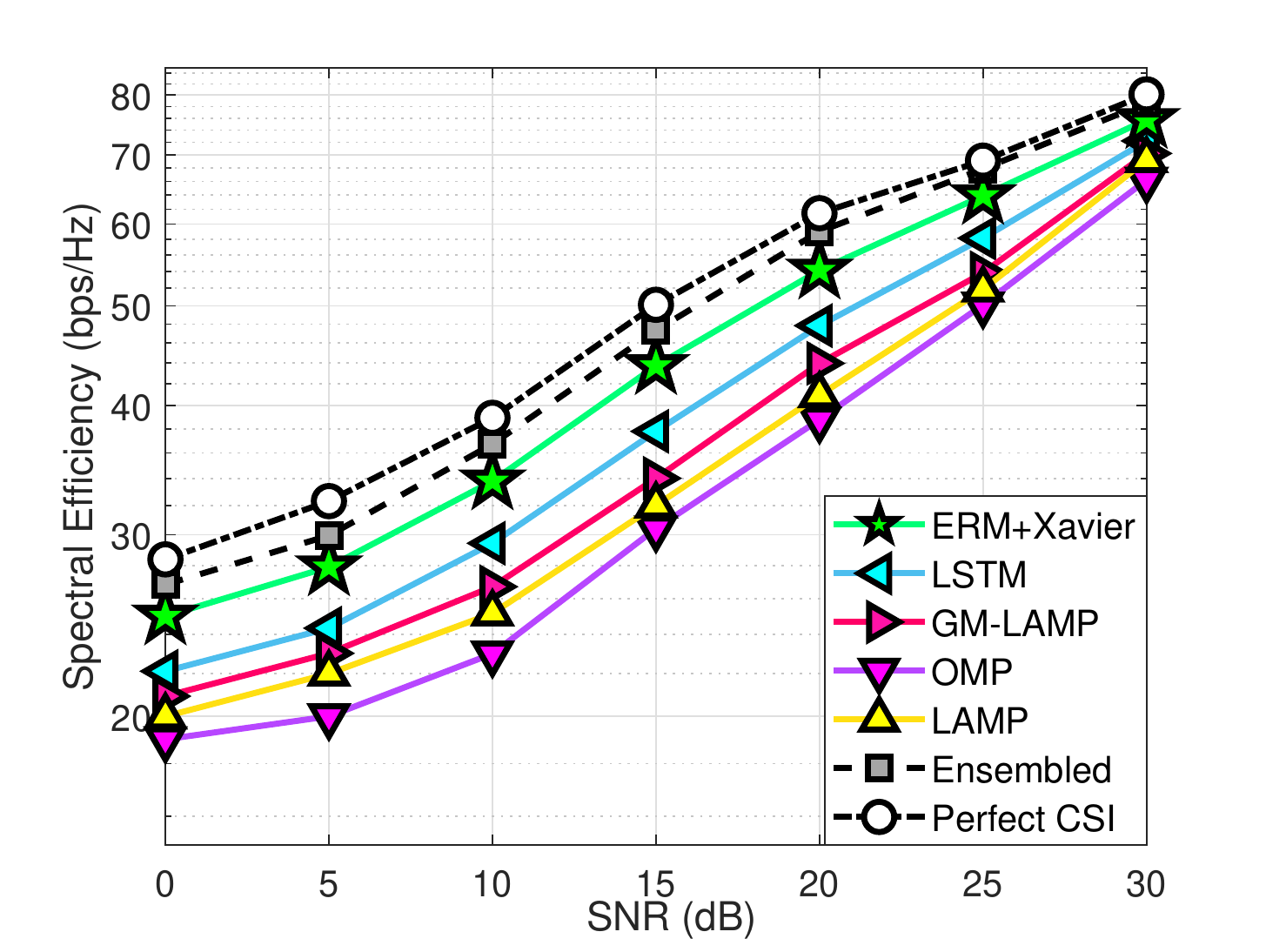}\caption{Average spectral efficiency against SNR.}
\label{fig:SE}
\end{figure}
The convergence is achieved in limited epochs thanks to the sparse connectivity in ERM architecture (specifically in reservoir layer(s)), as well as the utilization of Xavier initialization. One can compare the required epochs to converge in Fig.~\ref{fig:rmse}, to a typical ANN such as
spiking neural network (SNN) with far more epochs taking to converge (see Fig. 7 in \cite{SNN}). Fig.~\ref{fig:var} verifies that fine-tuning ERM with Xavier initializer technique makes the prediction variance lower, compared to the conventional ERM with random initial input weights, by up to 26\% and 49\%, for $\tau_{1}$ and $\tau_{1}$, respectively. In addition, compared to a single conventional ERM, the strong ensembled model leads to 31\% and 56\% reduction in prediction variance for $\tau_{1}$ and $\tau_{1}$, respectively.
\par  Fig.~\ref{fig:loss} shows the loss range upon training, validation and test phases for varying number of samples in a batch. Obviously, less errors are observed, when more number of samples are incorporated within a batch. What's more, the errors on validation and test phases are mostly dispersed in adjacent to the zero error line, which is negligible and verifies that ERM enjoys an accurate prediction.
Fig.~\ref{fig:SE} compares the achievable SE for the baseline schemes, where the ideal CSI is the upper bound to this end. According to Fig.~\ref{fig:nmse}, the proposed ERM+Xavier scheme enjoys a more accurate prediction among the counterparts, which accordingly results in 27\% and 13\% improvement in achievable SE of OMP\cite{OMP} and the LSTM\cite{LSTM1} schemes at SNR~=~15dB, respectively. By improving ERM+Xavier scheme one step further, the ensembled model is superior among all the schemes as well, with 36\% and 21\% more SE upon OMP\cite{OMP} and LSTM\cite{LSTM1} schemes at SNR~=~15dB, respectively. Whats more, as the SNR increases, all the schemes get closer to the perfect CSI, due to the better pilot transmission.
\section{Conclusions}
By tracking the historical features, we proposed a two-phase mmWave beamspace channel estimation scheme from a time-series prediction perspective (i.e., a mmWave beamspace channel tracking scheme). We fine-tuned the conventional ERM by means of Xavier initializer technique in the first phase, for lower prediction variance. The fine-tuned ERM shown to be capable of accurately learning from the past features and predicting the upcoming variations of the mmWave beamspace channel. In the second phase, the AdaBoost as an ensemble learning technique, applied on the fine-tuned Xavier-initialized ERM for even further reduced prediction variance, as well as an improved predicting precision. Simulations revealed a two-step improvement upon the counterparts in literature.
\section*{Acknowledgement}
This work is supported by the Academy of Finland: (a) ee-IoT n.319009, (b) EnergyNet n.321265/n.328869, and (c) FIREMAN n.326270/CHISTERA-17-BDSI-003; and by JAES Foundation via STREAM project.

\end{document}